\documentclass[twocolumn,showpacs,preprintnumbers,amsmath,amssymb]{revtex4}
\usepackage{epsfig}
\usepackage{amsmath}
\usepackage{graphicx}
\usepackage{amsfonts}
\usepackage{amssymb}
\usepackage{color}
\newcommand{\XXX}{}             
\newcommand{\YYY}{}             
\newcommand{\ZZZ}{}             
\newcommand{\BBB}{}             
\newcommand{\XVersion}{}

\newcommand{\beq}{\begin{equation}}
\newcommand{\eeq}{\end{equation}}

\newcommand{\eex}{E_{\rm exact}}
\newcommand{\eqlm}{E_{\rm QLM}}
\newcommand{\eqlmone}{E_{\rm QLM}^{(1)}}
\newcommand{\ewkb}{E_{\rm WKB}}
\newcommand{\Tfrac}[2]{#1/#2}

\begin{document}

\title{\BBB Quasilinearization Method and WKB\XVersion}

\author{R.~Krivec$^1$
       and
       V.~B.\ Mandelzweig$^2$}
\address{$^1$J. Stefan Institute, P.O.\ Box 3000, 1001 Ljubljana, Slovenia\\
         $^2$Racah Institute of Physics, Hebrew University, Jerusalem 91904, 
	 Israel}

\begin{abstract}
\smallskip Solutions obtained by the quasilinearization method (QLM) are
compared with the WKB solutions. While the WKB method generates an expansion
in powers of $\hbar$, the quasilinearization method (QLM) approaches the
solution of the nonlinear equation obtained by casting the Schr\"{o}dinger
equation into the {\XXX Riccati} form by approximating nonlinear terms by a
sequence of linear ones. It does not rely on the existence of any kind of
smallness parameter. It also, unlike the WKB, displays no unphysical turning
point singularities.  It is shown that both energies and wave functions
obtained in the first QLM iteration are accurate to a few parts of the
percent. Since the first QLM iterate is represented by the closed expression
it allows to estimate analytically and precisely the role of different
parameters, and influence of their variation on the properties of the
quantum systems. The next iterates display very fast quadratic convergence
so that accuracy {\BBB of energies and wave functions obtained after a few
iterations is extremely high, reaching 20 significant figures for the 
energy of the sixth iterate}. It is therefore demonstrated that the QLM
method could be preferable over the usual WKB method.
\end{abstract}

\maketitle

\section{Introduction} 

The quasilinearization method (QLM) was constructed as a generalization of
the Newton-Raphson method \cite{CB,RR} for the nonlinear differential
equations to yield rapid quadratic and often monotonic convergence to the 
exact solution. It was developed originally in theory of linear programming
by Bellman and Kalaba \cite{K,BK} to solve nonlinear ordinary and partial
differential equations and their systems. In the original works of Bellman
and Kalaba \cite{K,BK}, however, the convergence of the method has been
proven only under rather restrictive conditions of small intervals and
bounded, nonsingular forces \cite{VBM1} which generally are not fulfilled in
physical applications. This could explain an extremely sparse use of the
technique in physics, where only a few examples of the references to it
could be found \cite{C67,AIC87, RV, J, HR83}. Recently, however, it was
shown \cite{VBM1} by one of the present authors (VBM) that a different proof
of the convergence can be provided which allows to extend the applicability
of the method to realistic forces defined on infinite intervals with
possible singularities at certain points. This proof was generalized and
elaborated in the subsequent works \cite{VBM2, MT, KM1, KM2}.

In the first paper of the series \cite{VBM1}, the analytic results of the
quasilinearization approach were applied to the nonlinear Calogero equation
\cite{C67} in the variable phase approach to quantum mechanics, and the
results were compared with those of the perturbation theory and with the
exact solutions.  It was shown that the number of the exactly reproduced
perturbation terms doubles with each subsequent QLM iteration, which, of
course, is a direct consequence of a quadratic convergence.

The numerical calculation of higher QLM approximations to solutions of the
Calogero equation with different singular and nonsingular, attractive and
repulsive potentials performed in the work \cite{KM1} has shown that already
the first few iterations provide accurate and numerically stable answers for
any values of the coupling constant and that the number of iterations
necessary to reach a given precision increases only slowly with the coupling
strength. It was verified that the method provides accurate and stable
answers even for super singular potentials for which each term of the
perturbation theory diverges and the perturbation expansion consequently
does not exist.

In the third paper of the series \cite{MT} the quasilinearization method was
applied to other well known typical nonlinear ordinary differential
equations in physics, such as the Blasius, Duffing, Lane-Emden and
Thomas-Fermi equations which have been and still are extensively studied in
the literature. These equations, unlike the nonlinear Calogero equation 
\cite{C67} considered in references \cite{VBM1,KM1}, contain not only
quadratic nonlinear terms but various other forms of nonlinearity and not
only the first, but also higher derivatives. It was shown that again just a
small number of the QLM iterations yield fast convergent and uniformly
excellent and stable numerical results.

In the work \cite{KM2} the quasilinearization method was applied to quantum
mechanics by casting the Schr\"{o}dinger equation in the nonlinear Riccati
form and calculating the QLM approximations to bound state energies and wave
functions for a variety of potentials, most of which are not treatable with
the help of the perturbation theory or the $1/N$ expansion scheme. It was
shown that the convergence of the QLM expansion for both energies and wave
functions is very fast and that already the first few iterations yield
extremely precise results. In addition it was verified that the higher QLM
approximations, unlike those in $1/N$ expansion method, are not divergent at
any order.

The present work is devoted to comparison of QLM and WKB. Indeed, the
derivation of the WKB solution starts by casting the radial Schr\"{o}dinger
equation into nonlinear Riccati form and solving that equation by expansion
in powers of $\hbar$. It is interesting instead to solve this nonlinear
equation with the help of the quasilinearization technique and compare with
the WKB results. Such a procedure was performed in works \cite{RV,J}, where
it was shown that the first QLM iteration reproduces the structure of the
WKB series generating an infinite series of the WKB terms, but with
different coefficients. Besides being a better approximation, the first QLM
iteration is also expressible in a closed integral form. Similar conclusions
are reached for higher QLM approximations and it can be shown \cite{VBM3}
that the $p$-th QLM iteration yields the correct structure of the infinite
WKB series and reproduces $2^p$ terms of the expansion of the solution in
powers of $\hbar$ exactly, as well as a similar number of terms
approximately.

That the first QLM iteration already provides a much better approximation to
the exact solution than the usual WKB is obvious, not only from comparison
of terms of the QLM and WKB series \cite{RV, J, VBM3}, but also from the
fact that the quantization condition in the first QLM iteration leads to
exact energies for many potentials \cite{VBM3, KM3} such as for the Coulomb,
harmonic oscillator, P\"{o}schl-Teller, Hulthen, Hylleraas, Morse, Eckart
and some other well known physical potentials, which have a simple analytic
structure. By comparison, the WKB approximation reproduces exact energies
only in the case of the first two potentials.

The goal of this work is to point out that also for other potentials with
more complicated analytical structure QLM iterates provide much better
approximation than the usual WKB. If the initial QLM guess is properly
chosen the wave function in all QLM iterations, unlike the WKB wave
function, is free of unphysical turning point singularities.  Since the
first QLM iteration is given by an analytic expression
\cite{VBM1,VBM2,MT,KM1,RV,J}, it allows one to analytically estimate the
role of different parameters and the influence of their variation on
different characteristics of a quantum system. The next iterates display
very fast quadratic convergence so that accuracy of energies obtained after
a few iterations is extremely high, reaching up to 20 significant figures
for a sixth iterate as we show on the example of different widely used
physical potentials.

The paper is arranged as follows: in the second chapter we present the main
features of the quasilinearization approach to the solution of the
Schr\"{o}dinger equation, while in the third chapter we consider the
application of the method to computations with the anharmonic oscillator,
logarithmic, {\YYY two-power (double-well)}, and Wood-Saxon potentials and
to the two-body Dirac equation with static Coulomb potential. The final,
forth chapter is devoted to the discussion of the results, convergence
patterns, numerical stability, advantages of the method and its possible
future applications.

\section{Quasilinearization method} 

The usual WKB substitution 
\beq 
\chi(r)=C \exp\left(\lambda\int^r y(r') dr'\right)
\label{eq:ueq} 
\eeq 
converts the Schr\"{o}dinger equation to nonlinear Riccati form 
\beq 
\frac{dy(z)}{dz}+ (k^2(z)+y^2(z))=0. 
\label{eq:weq} 
\eeq 
Here {\XXX $k^2(z)=E-V-\Tfrac{l(l+1)}{z^2}$, $\lambda^2=\Tfrac{2 m}{\hbar^2}$}
and $z=\lambda r$.

The proper bound state boundary condition for potentials falling off at $z
\simeq z_0 \simeq \infty$ is $y(z)= \mathrm{const}$ at $z \geq z_0$. This
means that $y'(z_0) = 0$, so that Eq.\ (\ref{eq:weq}) at $z \simeq z_0$
reduces to $k(z_0)^2+y^2(z_0))=0$ or $y(z_0))= \pm i k(z_0)$. We choose here
to define the boundary condition with the plus sign, so that $y(z_0)= i
k(z_0)$.

The quasilinearization \cite{VBM1,MT,RV} of this equation gives a set of 
recurrence differential equations 
\beq
\frac{dy_{p}(z)}{dz}=y_{p-1}^2(z)-2 y_{p}(z)y_{p-1}(z)-k^2(z)
\label{eq:qeq} 
\eeq
with the boundary condition  $y_{p}(z_0)= i k(z_0)$.
 
The analytic solution \cite{RV} of these equations expresses the $p$-th
iterate $ y_{p}(z)$ in terms of the previous iterate:
\begin{eqnarray}
y_{p}(z)& =& f_{p-1}(z)- \nonumber \\
&&\XXX\int_{z_0}^{z}ds \frac{d\,f_{p-1}(s)}{ds}\;
\exp\left[-2 \int_{s}^{z} y_{p-1}(t) dt\right],\nonumber \\
 f_{p-1}(z)& =& 
\frac{y_{p-1}^2(z)-k^2(z)}{2 y_{p-1}(z)}.
\label{eq:ipqeq}
\end{eqnarray}
Indeed, differentiation of both parts of Eq.\ (\ref{eq:ipqeq}) leads
immediately to Eq.\ (\ref{eq:qeq}) which proves that $y_{p}(z)$ is a
solution of this equation. The boundary condition is obviously satisfied
automatically.

To utilize the recurrence relation (\ref{eq:ipqeq}) for wave function
computation one has to pick up a proper initial guess. For the zeroth
iterate $y_{0}(z)$ it seems natural to choose the zero WKB approximation
that is to set $y_{0}(z)=i k(z)$, which in addition automatically satisfies
the boundary condition. However, one has to be aware that this choice has
unphysical turning point singularities. According to the existence theorem
for linear differential equations \cite{In}, if $y_{p-1}(z)$ in Eq.\
(\ref{eq:qeq}) is a discontinuous function of $z$ in a certain interval,
then $y_{p}(z)$ or its derivatives may also be discontinuous functions in
this interval, so consequently the turning point singularities of $y_{0}(z)$
may propagate to the next iterates. To avoid this we choose \cite{KMT} the
Langer WKB wave function \cite{Lan} as the zero iteration. This function
near the turning points $a$ and $b$ is given by the simple analytic
expression \cite{ftn}
\begin{eqnarray}
\chi_i(r)&=&c_i \sqrt{\frac{S_i^{\frac{1}{3}}(r)}{\left|k(r)\right|}}
\mathrm{Ai} \left[d\; S_i^{\frac{1}{3}}(r)\right],\nonumber
\\ S_i(r)&=&\frac{3}{2} \lambda
\left| \int_{i}^{r} \left| k(s)\right| ds \right|.
\label{eq:lan}
\end{eqnarray} 
Here Ai denotes the Airy function, $i=a,b$, {\ZZZ $k^2(r)=2 m (E-V(r)) -
\Tfrac{(l+\Tfrac{1}{2})^2}{r^2}$}, $d$ is $-1$ for $a<r<b$, and $1$ for $r
\leq a$, $r \geq b$, and $c_a=1$, $c_b=(-1)^n$, where $n=0,1,2,\XXX {\ldots}
$ is the number of the bound state. {\ZZZ $\chi_a(r)$ and $\chi_b(r)$ are
continuous across the turning points and coincide with the usual WKB
solution far from them. It is easy to check that $\chi_a(r)$ and $\chi_b(r)$
coincide at some point in the interval $(a,b)$ between the turning points,
and that their values, but not derivatives, can be matched at that point.}

\section{Examples}

To show that the first QLM iteration gives very accurate results for wave
functions and energies, as well as demonstrate very fast convergence of the
next iterates let us consider five typical examples of potentials of rather
different form used in atomic, nuclear and quark physics. 

Let us start from the anharmonic oscillator $V(r)=\frac{1}{2}r^5$. This
potential is typically used in different nuclear, quark and quantum field
theory models. The exact energy of the ground state of this oscillator is
{\XXX 2.044\;579\;657\;447\;355\;635\;36} in atomic units with mass set to
unity, $m = 1$. This result is obtained {\XXX by us by a calculation using 
the Runge-Kutta method in quadruple} precision. The WKB energy is different
by 4.5\% and equals 1.95159 in the same units, while the first-iteration QLM
energy equals 2.04528 and differs from the exact energy only by 0.034\%. The
QLM energy coincides with the exact energy in all twenty digits after the
sixth iteration. 

For the first excited state the exact energy is {\XXX
6.713\;546\;501\;445\;253\;110\;53}, while the WKB and first-iteration QLM
energies are 6.656623 and 6.713952 and are different from the exact energy
by 0.84\% and 0.006\% respectively.  The similar picture exists for the 
second excited state where the exact energy is {\XXX
12.767\;866\;541\;180\;535\;228\;88}. The WKB and first-iteration QLM
energies are {\XXX 12.72396} and 12.76796 and are different from the exact
energy by 0.34\% and 0.0007\% respectively. Again, for both first and second
excited states the QLM energies differ from the exact energies only in the
twentieth digit after the sixth iteration.

\begin{figure}
\begin{center}
\epsfig{file=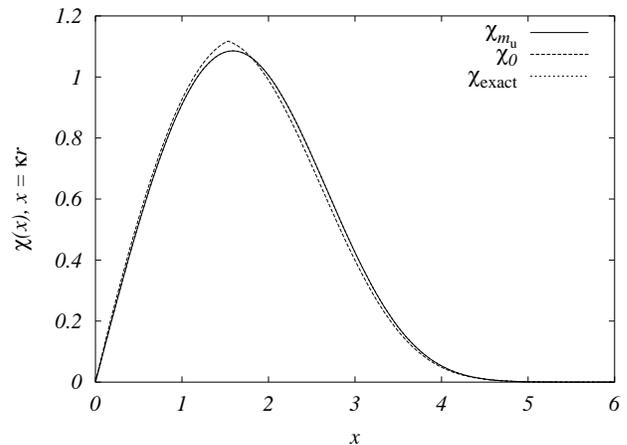,width=87mm}
\end{center}
\caption{Comparison of the Langer WKB solution $\chi_0$ (dashed curve), the
exact solution $\chi_{\rm exact}$ (dotted curve) and the first QLM iterate
$\chi_{m_u}$ (solid curve) for the ground state of the anharmonic oscillator
$V(r)=\frac{1}{2}r^5$.  The last two are indistinguishable on the plot. Here
$x = \kappa r$, $\kappa^2 = 2mE/\hbar^2$.} \label{fig1}
\end{figure}

The graphs corresponding to the Langer WKB solution, the exact solution and
the first QLM iterate for the ground state are displayed in Fig.\
\ref{fig1}. One can see that while the Langer solution is noticeably
different from the exact solution, the curve of the first QLM iteration is
indistinguishable from the exact curve.

\begin{figure}
\begin{center}
\epsfig{file=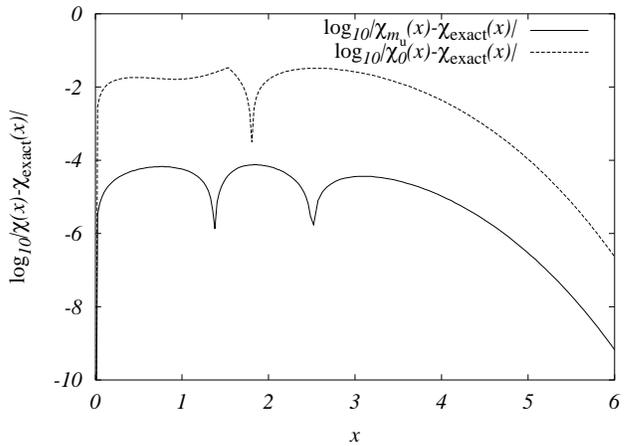,width=87mm}
\end{center}
\caption{Logarithm of the difference between the exact $\chi_{\rm exact}$
and WKB solutions $\chi_{0}$ (dashed curve) and between the exact solution
and the first QLM iterate $\chi_{m_u}$ (solid curve) for the ground state of
the anharmonic oscillator. {\XXX The dips} on the graphs are artifacts of
the logarithmic scale, since the logarithm of the absolute value of the
difference of two solutions goes to minus infinity at points where the
difference changes sign. The overall accuracy of the solution can be
inferred only at $x$ values not too close to the {\XXX dips}.} \label{fig2}
\end{figure}

This could be followed more precisely by looking at Fig.\ \ref{fig2} where
the logarithm of the difference between the exact and WKB solutions and
between the exact solution and the first QLM iteration are shown. One can
see that the difference between the exact solution and the first QLM
iteration is two orders of magnitude smaller than the difference between the
exact and the WKB solutions, that is one QLM iteration increases the
accuracy of the result by two orders of magnitude. Note that the {\XXX dips}
on the graphs are artifacts of the logarithmic scale, since the logarithm of
the absolute value of the difference of two solutions goes to minus infinity
at points where the difference changes sign. The overall accuracy of the
solution can be inferred only at $x$ values not too close to the {\XXX
dips}. 

\begin{figure}
\begin{center}
\epsfig{file=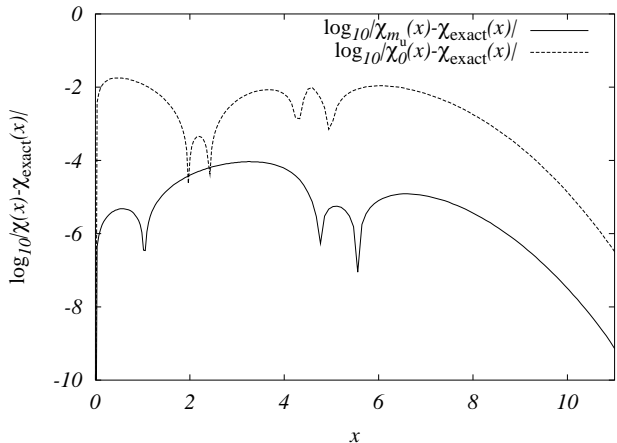,width=87mm}
\end{center}
\caption{As in Fig.\ \ref{fig2}, but for the excited state of the
oscillator potential.}
\label{fig3}
\end{figure}

The accuracy of the WKB approximation increases for higher excitations.
Therefore in the case of the excited state both the Langer WKB and QLM
curves are indistinguishable from the exact one. Fig. \ref{fig3} show,
however, that also in this case the difference between the exact solution
and the first QLM iteration is by two orders of magnitude smaller than the
difference between the exact and the WKB solutions.

Another interesting example is the modified Coulomb potential 
\beq \XXX
V(r)= - \frac{1}{2 \rho} + \frac{l (l+1) - \frac{1}{4} 
\alpha^2}{\rho^2} + 
\frac{\frac{3}{4} \alpha^2}{\rho^2 (\rho +\alpha^2)^2},\; 
\rho = \alpha E r  \nonumber
\eeq
which is obtained when the equal masses two-body Dirac equation with the
static Coulomb interaction is reduced to the Schr\"{o}dinger equation
\cite{M, S}. The exact energy of the ground state with quantum numbers 
{\XXX $(N,L,S,J)=(1,0,0,0)$ is 0.999\;993\;340\;148\;538\;880\;1} in atomic
units with double mass set to unity, {\XXX $2M = 1$}. This result was
obtained in the work \cite{S} by an elaborate computation using the finite
element method and confirmed {\XXX by ourselves using the Runge-Kutta method
in quadruple} precision.  The WKB energy equals {\XXX 0.999\;986\;680 and
differs} from the exact one by {\XXX $6.6\times10^{-4}$}. The
first-iteration QLM energy equals {\XXX 0.999\;993\;335} and differs from 
the exact one only by {\XXX $5\times 10^{-7}$}.  The QLM energy coincides
with the exact one in all given digits after the sixth iteration.

\begin{figure}
\begin{center}
\epsfig{file=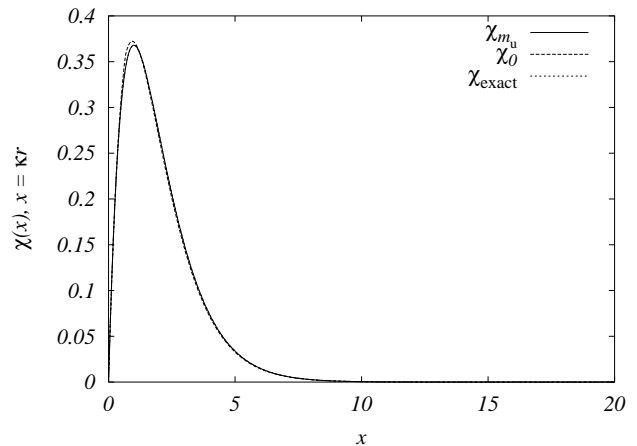,width=87mm}
\end{center}
\caption{As in Fig.\ \ref{fig1}, but for the ground state 
with quantum numbers {\XXX $(N,L,S,J)=(0,0,0,0)$}
in the
modified Coulomb potential {\XXX
$V(r)= - \frac{1}{2 \rho} + 
\Tfrac{\left(l (l+1) - \frac{1}{4}\alpha^2\right)}{\rho^2} + 
\Tfrac{\frac{3}{4} \alpha^2}{\rho^2 (\rho +\alpha^2)^2}$},
$\rho = \alpha E r$.}
\label{fig1a}
\end{figure}

\begin{figure}
\begin{center}
\epsfig{file=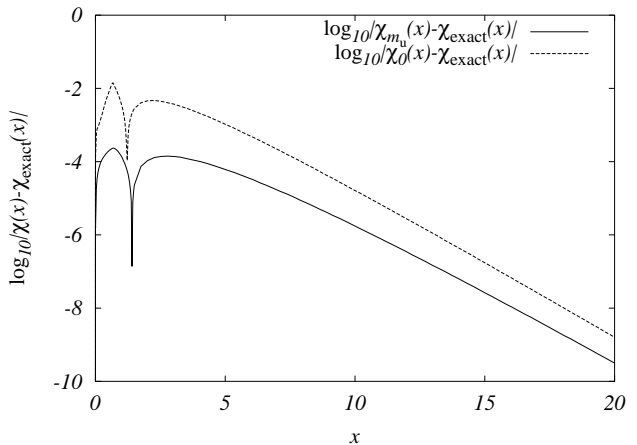,width=87mm}
\end{center}
\caption{As in Fig.\ \ref{fig2}, but for the ground state in the
modified Coulomb potential.}
\label{fig2a}
\end{figure}

The {\XXX graph} Fig.\ \ref{fig1a} of the exact, WKB and QLM {\XXX ground}
state wave functions is similar to Fig.\ \ref{fig2}.

The graph Fig.\ \ref{fig2a} for their differences for this case is similar
to Fig.\ \ref{fig2} and shows that the difference between the exact wave
function and the first QLM iteration is by two orders of magnitude smaller
than the difference between the exact and the WKB solutions. Thus also in
this case just one QLM iteration increases the accuracy of the wave function
by a remarkable two orders of magnitude.

\begin{figure}
\begin{center}
\epsfig{file=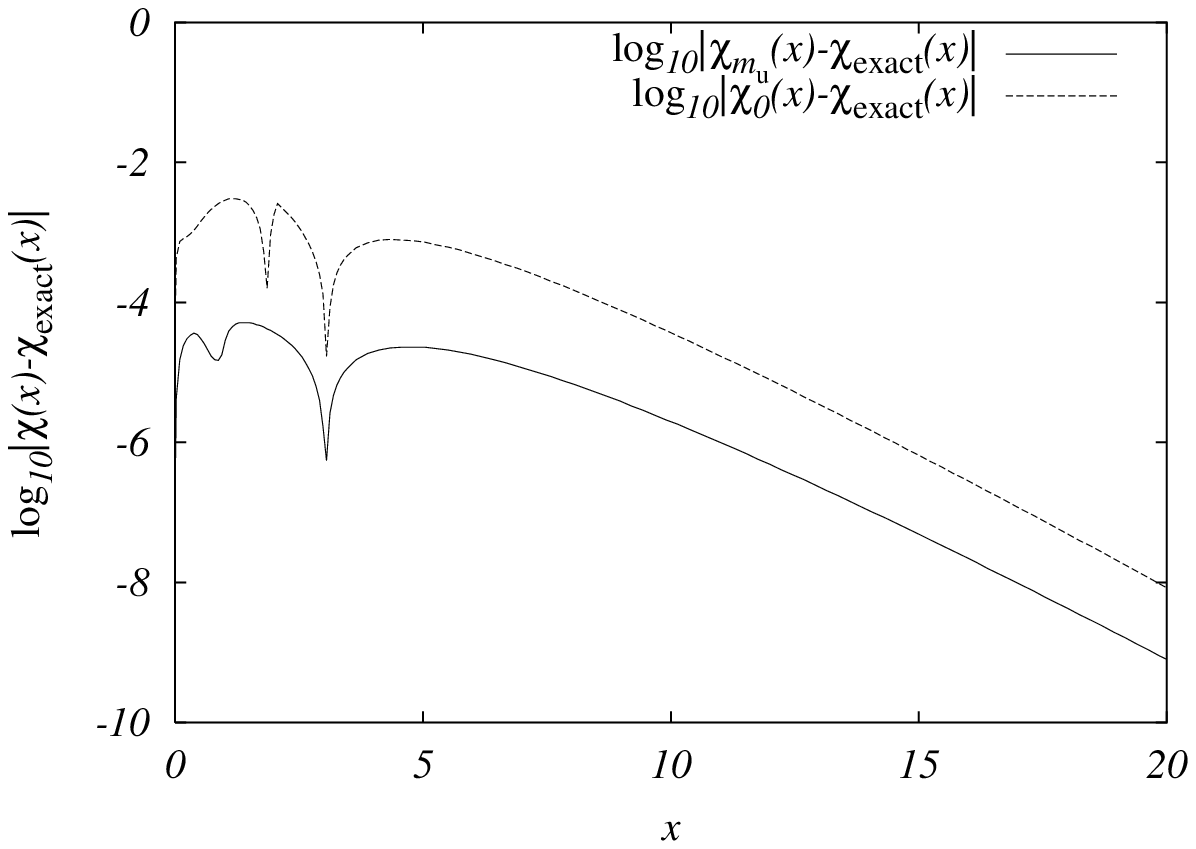,width=87mm}
\end{center}
\caption{As in Fig.\ \ref{fig2}, but for the excited state with
quantum numbers {\XXX $(N,L,S,J)=(2,0,0,0)$} in the
modified Coulomb potential.}
\label{fig1b}
\end{figure}

\begin{figure}
\begin{center}
\epsfig{file=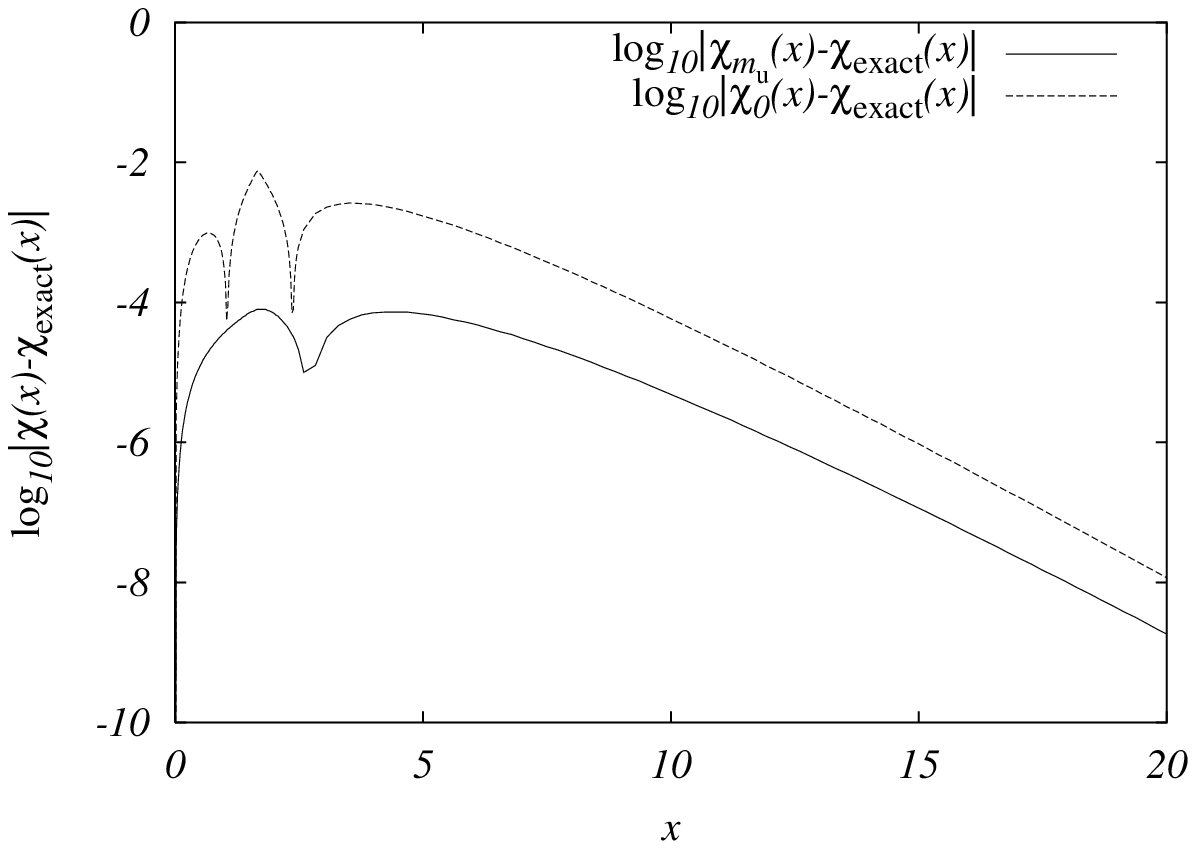,width=87mm}
\end{center}
\caption{As in Fig.\ \ref{fig2}, but for the excited state with
quantum numbers {\XXX $(N,L,S,J)=(1,1,0,1)$} in the modified 
Coulomb potential.}
\label{fig2b}
\end{figure}

\begin{figure}
\begin{center}
\epsfig{file=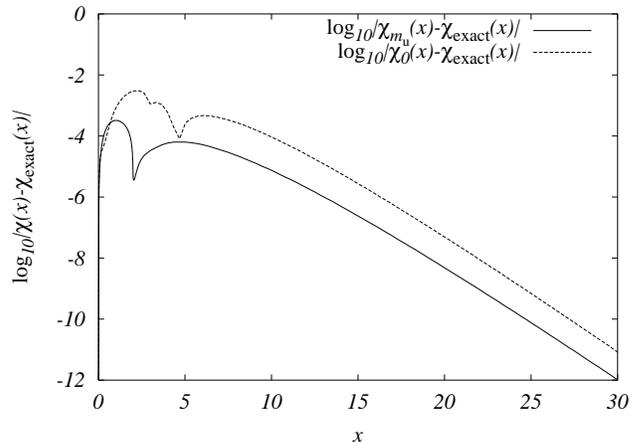,width=87mm}
\end{center}
\caption{As in Fig.\ \ref{fig2}, but for for the excited state with
quantum numbers {\XXX $(N,L,S,J)=(2,1,0,1)$} in the
modified Coulomb potential.}
\label{fig3b}
\end{figure}

The results for the ground and excited states with different quantum numbers
{\XXX $(N,L,S,J)$ for the modified Coulomb potential are summed up} in Table
1 and also in Figs.\ \ref{fig1b}, \ref{fig2b}, \ref{fig3b} where the the
differences between the exact wave function and the first QLM iteration and
between the exact and the WKB solutions are displayed. We see, that though
the accuracy of the WKB approximation increases for excited states and
states with higher orbital momenta, also in these cases one QLM iteration
increases the accuracy of the wave function by at least two orders of 
magnitude.

\begin{widetext}

\begin{table}[t]
\begin{center}
\caption{%
The WKB, first iteration QLM, full QLM and exact binding energies for 
different potentials. $K$ denotes the number of QLM iterations, $m$ denotes
the (reduced) mass of the particle; $D_1 = 10^2 (\eex - \ewkb)/\eex$, and
$D_2 = 10^2 (\eex - \eqlmone)/\eex$. (In the first entry, $D_1$ and $D_2$
are given in the form $x[y]$ which stands for $x\times10^y$.) {\ZZZ $\ewkb$
and $\eqlmone$ are given to a limited number of digits, since $\eqlmone$ is
slightly dependent on the discontinuity in the derivative of the Langer WKB
solution at its joining point between the turning points.} {\BBB This
dependence disappears in higher QLM iterations. For the high-precision
results, the number of correct digits is at least 18, and $E$ tends to be
slightly more precise. Consequently we give the numbers to 20 or 21 digits.}
Since the computer arithmetic was quadruple precision (128-bit, about 30
decimal places), the differences in the last digits of $\eqlm$ and $\eex$
reflect the different methods used. The state is labeled by $nl$ except for
the Breit-Coulomb problem where the labels are $N, L, S, J$.  The plus sign
for the {\YYY two-power} potential stands for the ground (symmetric) state
in the corresponding one-dimensional {\YYY double-well} potential; the minus
sign stands for the regular states of the two-power potential or the
antisymmetric state of the corresponding one-dimensional {\YYY double-well}
potential, respectively. }
\begin{tabular}{|c|c|c|c|rrr|r|cc|}
\hline
Potential                                & $m$   & State &   $\ewkb$      &  $\eqlmone$    &  $\eqlm$                 & $K$& $\eex$                  & $D_1$  & $D_2$ \\
\hline
Breit-                                & 1 & 1\,0\,0\,0   & 0.999986679987 & 0.999993335480 &   0.99999334014853888012 &  6 &   0.99999334014853888016 &  7[-4]  & ~5[-7] \\
Coulomb                               &   & 2\,0\,0\,0   & 0.999996670008 & 0.999998335239 &   0.99999833502466540218 &  7 &   0.99999833502466540223 &  2[-4]  & -2[-8] \\
                                      &   & 1\,1\,0\,1   & 0.999996670037 & 0.999998335831 &   0.99999833501727839123 & 44 &   0.99999833501727839122 &  2[-4]  & -8[-8] \\
                                      &   & 2\,1\,0\,1   & 0.999998520016 & 0.999999260060 &   0.99999926000774772931 & 47 &   0.99999926000774772931 &  7[-5]  & -1[-8] \\
&&&&&&&&& \\                                                                                                                                                              
$\log r$                          &$\frac{1}{2}$ &  $1s$ &  1.05346726985 &   1.044738     &   1.04433226746060809298 &  5 &   1.04433226746060809380 & -0.88  & -0.039 \\
                                         &       &  $2s$ &  1.850802588   &   1.8475       &   1.84744258030447816386 &  5 &   1.84744258030447816385 & -0.18  & -0.003 \\
                                         &       &  $3s$ &  2.299218712   &   2.289659     &   2.28961571419653762102 &  5 &   2.28961571419653762102 & -0.42  & -0.002 \\
&&&&&&&&& \\                                                                                                                                                              
$\frac{1}{2}r^5$                         & 1     & $1s$  &  1.9515942     &  2.045279      &   2.04457965744735563534 &  6 &   2.04457965744735563536 &  4.5   & -0.03  \\
                                         &       & $2s$  &  6.656623      &  6.713952      &   6.71354650144525311020 &  6 &   6.71354650144525311053 &  0.85  & -0.006 \\
                                         &       & $3s$  & 12.72396~      & 12.76796~      &  12.7678665411805352297~ &  6 &  12.7678665411805352289~ &  0.34  & -0.001 \\
&&&&&&&&& \\                                                                                                                                                              
\XXX $\frac{-24}{1+\exp{\frac{r-1}{0.2}}}$  & {\XXX 1}  
                                                 &  $1s$ & -17.61192~~    & -17.5432~      & -17.5597967410317970585~ &  5 & -17.5597967410317970589~ & -0.30  &  0.095 \\
                                         &       &  $2s$ &  -7.190505     &  -7.37920      &  -7.37854164337449079226 &  5 &  -7.37854164337449079262 &  2.5   & -0.009 \\
                                         &       &  $3s$ &  -0.029269     &  -0.105156     &  -0.10819568493119384889 &  6 &  -0.10819568493119384933 & 72.9   &  2.8   \\
&&&&&&&&& \\                                                                                                                                                              
$\frac{1}{2}g^2{(r^2-a^2)^2}   $         & 1     &  $1s$+&   0.484067
\footnote{\BBB Includes the tunneling correction to $E$.}      
                                                                            &  0.483017 \footnote{\BBB Initial WKB approximation includes tunneling correction to $E$.}      
                                                                                           &   0.48295865991331554844   &  6 &   0.48295865991331554820   & -0.98  & -0.009 \\
&&&&&&&&& \\                                                                                                                                                              
                                         &       & $1s-$ & 0.49734197     & 0.484218       &   0.48314820684089227025 &  6 &   0.48314820684089227025 & -2.9   & -0.22  \\
                                         &       & $2s-$ & 1.39372888     & 1.373747       &   1.37363583606219407956 &  6 &   1.37363583606219407958 & -1.5   & -0.008 \\
                                         &       & $3s-$ & 2.17217337     & 2.178319       &   2.17745782251542955262 &  6 &   2.17745782251542955243 &  0.24  & -0.040 \\
\hline
\end{tabular}
\label{qlm2tab01}
\end{center}
\end{table}
\end{widetext}

\begin{figure}
\begin{center}
\epsfig{file=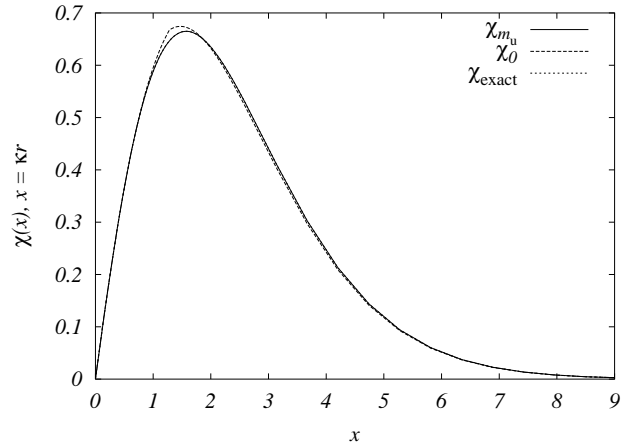,width=87mm}
\end{center}
\caption{As in Fig.\ \ref{fig1}, but for the ground state of the
logarithmic potential $V = log(r)$.}
\label{fig4}
\end{figure}

\begin{figure}
\begin{center}
\epsfig{file=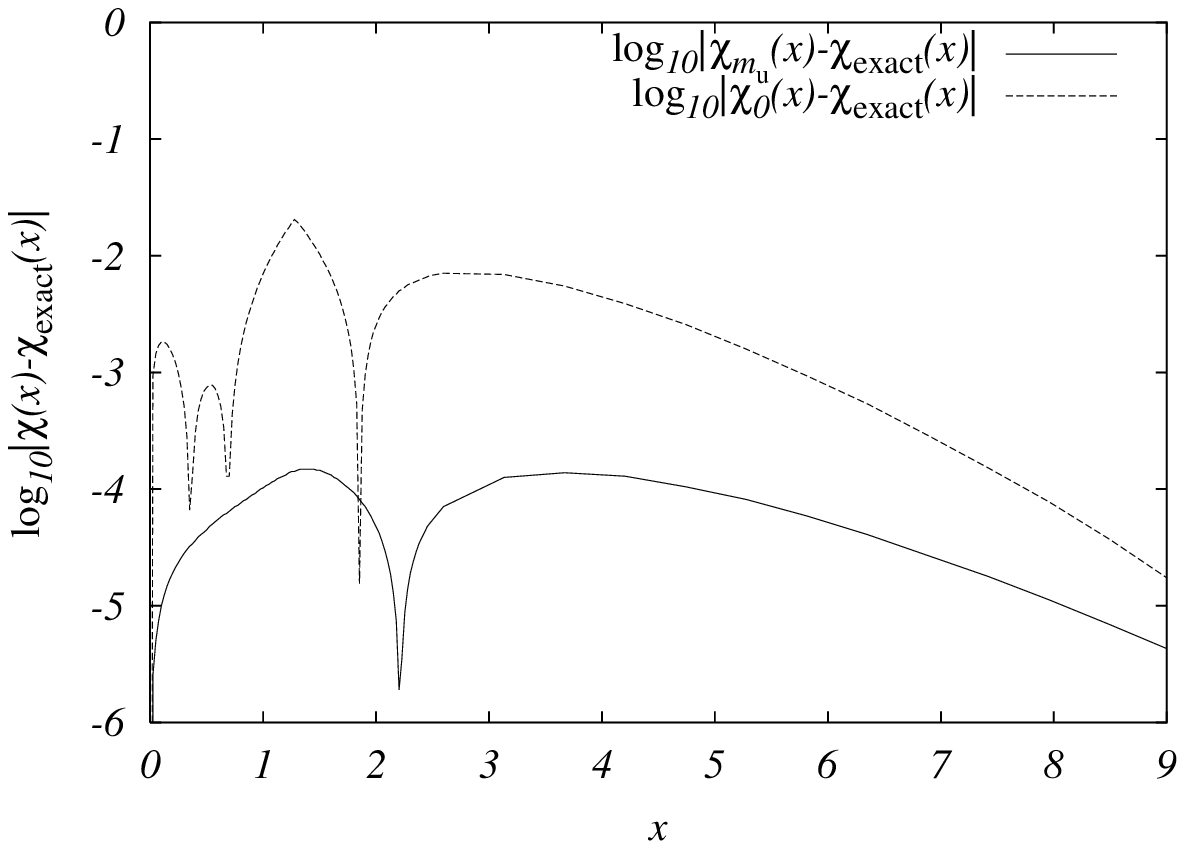,width=87mm}
\end{center}
\caption{As in Fig.\ \ref{fig2}, but for the ground state of the
logarithmic potential.}
\label{fig5}
\end{figure}

\begin{figure}
\begin{center}
\epsfig{file=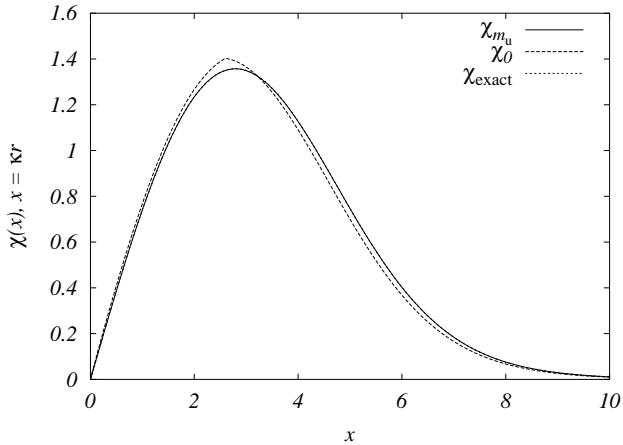,width=87mm}
\end{center}
\caption{As in Fig.\ \ref{fig1}, but for the ground state of {\XXX the}
Wood-Saxon potential {\XXX $V = - V_0 / (1 + \exp((r - R)/a))$}, with 
{\XXX $V_0 = 24$, $R = 1$, $a = 0.2$}.}
\label{fig6}
\end{figure}

\begin{figure}
\begin{center}
\epsfig{file=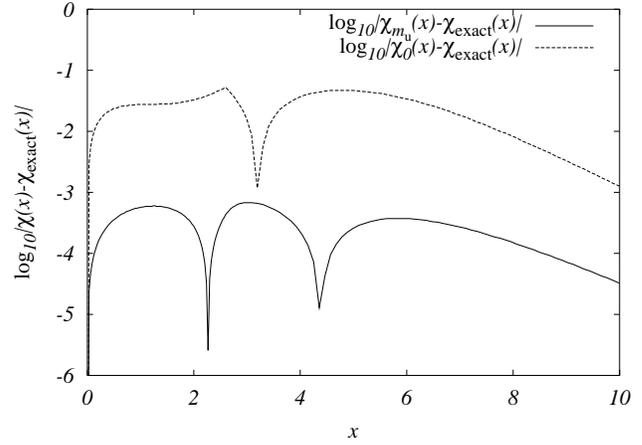,width=87mm}
\end{center}
\caption{As in Fig.\ \ref{fig2}, but for the ground state of {\XXX the}
Wood-Saxon potential.}
\label{fig7}
\end{figure}

\begin{figure}
\begin{center}
\epsfig{file=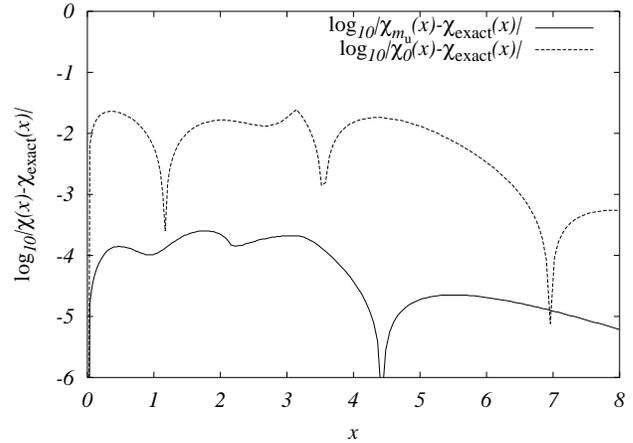,width=87mm}
\end{center}
\caption{As in Fig.\ \ref{fig2}, but for the first excited state of {\XXX the}
Wood-Saxon potential.}
\label{fig8}
\end{figure}

\begin{figure}
\begin{center}
\epsfig{file=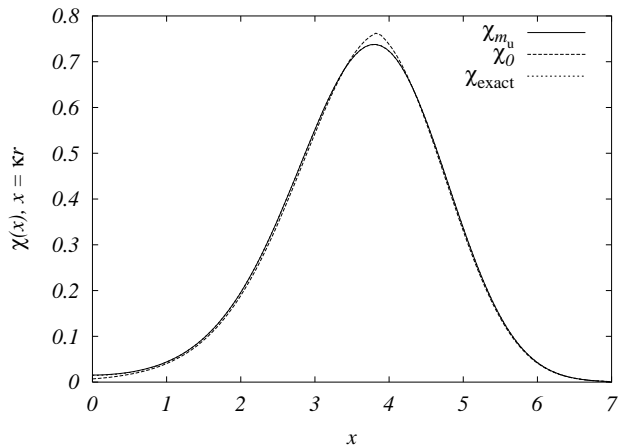,width=87mm}
\end{center}
\caption{As in Fig.\ \ref{fig1}, but for the ground (symmetric) state of the
{\YYY double-well potential $V = \frac {1}{2} g^2 (r^2 - a^2)^2$,
$g^2={1}/{4a^2}$, $a=4$ in one dimension}.} \label{fig9}
\end{figure}

\begin{figure}
\begin{center}
\epsfig{file=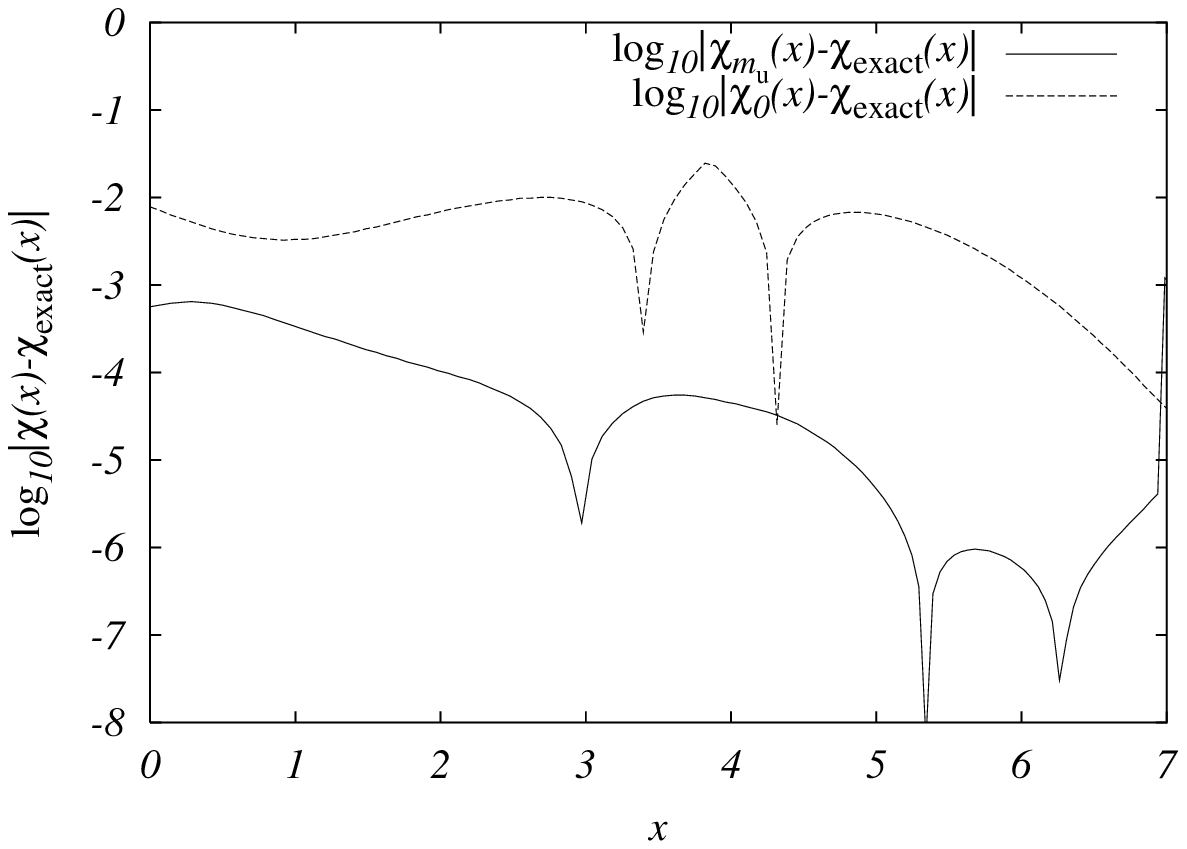,width=87mm}
\end{center}
\caption{As in Fig.\ \ref{fig2}, but for the state of Fig.\ \ref{fig9}.}
\label{fig10}
\end{figure}

\begin{figure}
\begin{center}
\epsfig{file=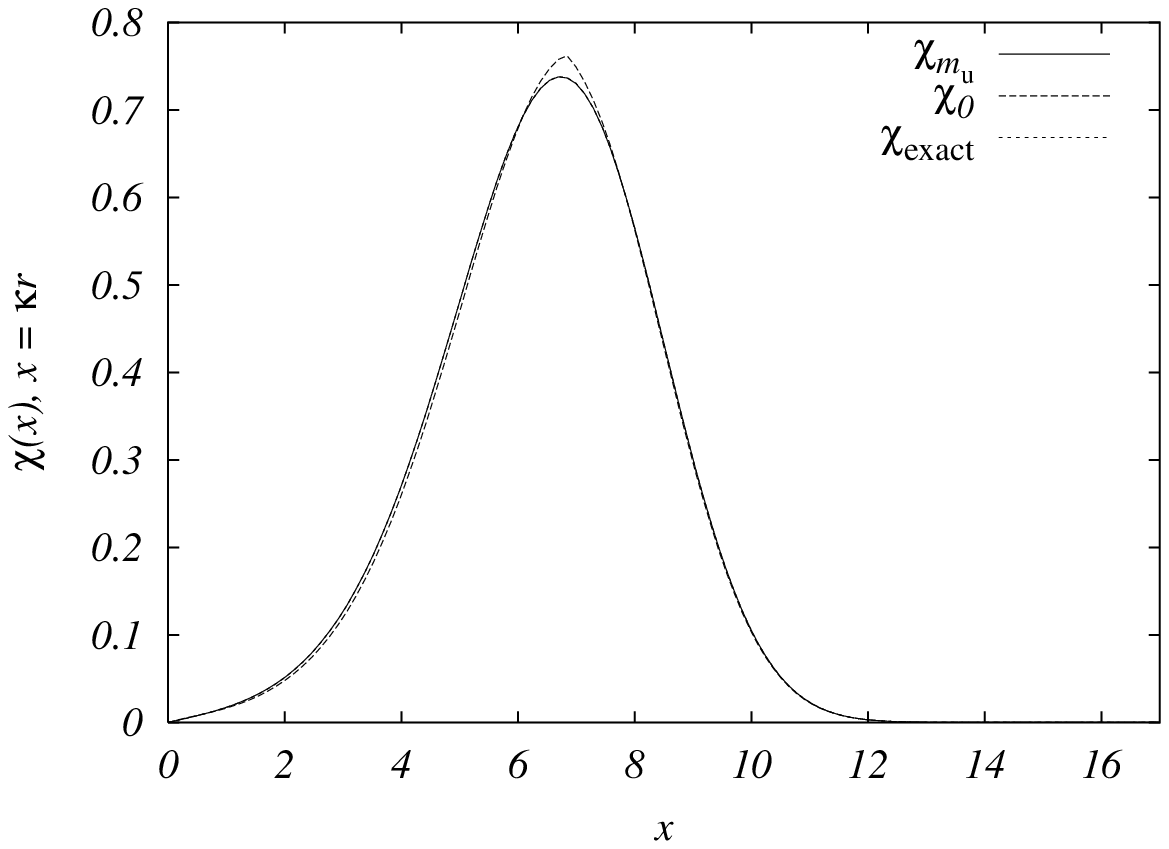,width=87mm}
\end{center}
\caption{As in Fig.\ \ref{fig1}, but for the ground state of the {\YYY
two-power} potential $V = \frac {1}{2} g^2 (r^2 - a^2)^2$, $g^2={1}/{4a^2}$,
$a=4$., or for the first (antisymmetric) excited state of the corresponding
{\YYY double-well potential in one dimension}.} \label{fig11}
\end{figure}

\begin{figure}
\begin{center}
\epsfig{file=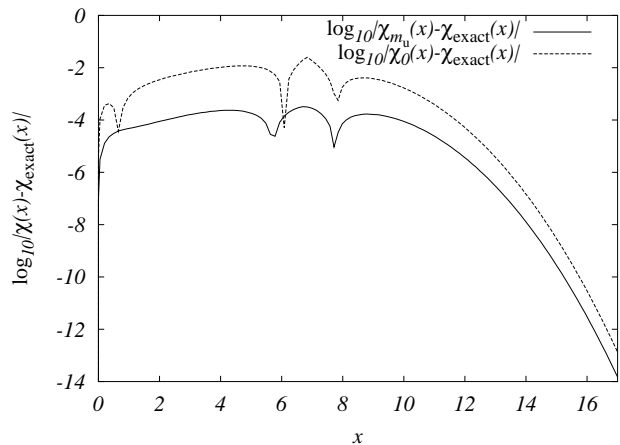,width=87mm}
\end{center}
\caption{As in Fig.\ \ref{fig2}, but for the state of Fig.\ \ref{fig11}.}
\label{fig12}
\end{figure}

The other examples considered in this paper are the logarithmic $V(r)=\log
r$, Wood-Saxon {\XXX $V = - V_0 / (1 + \exp((r - R)/a))$} and the {\YYY
two-power (double-well)} $V(r)=\frac {1}{2} g^2 (r^2 - a^2)^2$ potentials,
the results for which are {\XXX summarized} in Table 1.  {\XXX The graphs
corresponding to different states of these potentials are shown in Figs.\
\ref{fig4} - \ref{fig12}.} The first two potentials are used respectively
for computations in quark and nuclear physics. The {\XXX double-well}
potential, that is the quartic potential in one dimension with degenerate
minima, is typically {\XXX studied in quantum field theory and in the
framework of the} tunneling problem in quantum mechanics. Its perturbation
series does not converge and different alternative nonperturbative
approaches are therefore explored since the description of tunneling between
two minima should be necessarily nonperturbative (see, for example,
reference \cite{Lee} and the references therein). 

{\YYY In particular, in the paper \cite{BB} using the $1/N$ expansion
method, the tunneling terms were not included for the symmetric (ground)
state of the {\YYY double well} potential in one dimension, giving the $1/N$
energy of $0.48305$ compared to the exact energy, $0.48295{\ldots}$ In
addition, in our calculation it is easy to specify the boundary condition at
$r=0$ in this particular case (where $\chi(0) \neq 0$), so we can calculate
on the interval $r \ge 0$ only: because we do the QLM iteration on the
function $u(\kappa r)=\arctan\bigl(-\Tfrac{\kappa\chi(r)}{\chi'(r)}\bigr)$,
we have simply $u(0)=-\frac{\pi}{2}$.} {\ZZZ This can easily be seen by
taking into account that $\chi(r)$ has an even-power Taylor expansion at
$r=0$.} {\BBB We use the tunneling term just to correct the energy of the
initial WKB approximation, changing the usual WKB quantization condition to
\begin{equation}
\int_a^b k(r) dr=(n+\frac{1}{2})\pi - \frac{1}{2}e^{-\int_{-a}^a K(r) dr} \nonumber
\end{equation}
where the second term on the RHS is the tunneling term \cite{Park}; $k(r) =
iK(r)$ and $n=0,1,2,{\ldots} $ is the number of the bound state. The
tunneling correction affects the 1st QLM iteration but of course not the
full QLM calculation, where the boundary conditions completely specify the
converged solution.}

\section{Conclusion}

One can show \cite{VBM3, KM3} that the approximation by the first QLM
iterate in Eq.\ (\ref{eq:ipqeq}) leads to exact energies for many well 
known physical potentials {\XXX such} as the Coulomb, harmonic oscillator,
P\"{o}schl-Teller, Hulthen, {\XXX Hylleraas}, Morse, Eckart, etc.  For other
potentials which have more complicated analytical structure we show on
examples of the anharmonic oscillator, logarithmic, {\YYY two-power
(double-well)}, and Wood-Saxon potentials and for the solution of the
two-body Dirac equation with static Coulomb potential, that the use of the
Langer WKB wave function as an initial guess already in the first QLM
approximation gives energies and wave functions two orders of magnitude more
accurate than the WKB results. Such a QLM solution, unlike the usual WKB
solution, displays no unphysical turning point singularities. Since the
first QLM iterate is given by an analytic expression (\ref{eq:ipqeq}) for
$p=1$ it allows one to estimate analytically the role of different
parameters and their influence on properties of a quantum system with much
higher precision than provided by the WKB approximation.  In addition, it
was shown that six QLM iterations are typically enough to obtain both the
wave function and energy with the {\BBB accuracy of twenty significant
digits.}

\section*{Acknowledgments} 

The research was supported by the Bilateral
Cooperation Program at the Ministry of education, science and sport of
Slovenia (RK) and by the Israeli Science Foundation grant 131/00 (VBM).


\begin{references}

\bibitem{CB}
{Samuel D. Conte  and Carl de Boor, {\it Elementary numerical analysis},
 McGraw Hill International Editions,1981.}

\bibitem{RR}
{Anthony Ralston  and Philip Rabinowitz, {\it A first course in 
numerical analysis},
 McGraw Hill International Editions, 1988.}

\bibitem{K}
{R.~Kalaba, J.\ Math.\ Mech.\ {8}, 519 (1959).}

\bibitem{BK} {R.~E.\ Bellman
and R.~E.\ Kalaba, {\it Quasilinearization and Nonlinear
Boundary-Value Problems}, Elsevier Publishing Company, New York,
1975.}

\bibitem{C67} {F.~Calogero, {\it Variable Phase Approach to Potential
Scattering}, Academic Press, New York, 1975.}

\bibitem{AIC87}  {A.~A.\ Adrianov, M.~I.\ Ioffe and F.~Cannata,
Modern Phys.\ Lett.\ {11}, 1417 (1987).} 

\bibitem{RV}  {K.~Raghunathan and R.~Vasudevan, J.\ Physics A:
Math.\ Gen.\ {20}, 839 (1987).} 

\bibitem{J}  {M.~Jameel, J.~Physics A: Math.\ Gen.\ {21}, 1719
(1988).} 

\bibitem{HR83}  {M.~A.\ Hooshyar and M.~Razavy, Nuovo Cimento
{B75}, 65 (1983).} 

\bibitem{VBM1}
{V.~B.\ Mandelzweig, J.\ Math.\ Phys.\ {40}, 6266 (1999).} 

\bibitem{VBM2}
{V.~B.\ Mandelzweig, Few-Body Systems Suppl.\ {14}, 185 (2003).}

\bibitem{KM1}
{R.~Krivec and V.~B.\ Mandelzweig, Computer Physics
 Comm.\ {138}, 69 (2001).}

\bibitem{MT}  { V.~B.\ Mandelzweig and F.~Tabakin, Computer 
Physics Comm.\ {141}, 268 (2001).}

\bibitem{KM2}
{R.~Krivec and V.~B.\ Mandelzweig, Computer Physics
 Comm.\ {152}, 165 (2003).}

\bibitem{VBM3}
{V.~B.\ Mandelzweig, {\it Quasilinear approach to summation of the 
WKB series}, 2004, submitted for publication.}

\bibitem{KMT}
R.~Krivec, V.~B.\ Mandelzweig and F.~Tabakin,
{\BBB Few-Body Systems 34, 57 (2004).}

\bibitem{KM3}
{R.~Krivec and V.~B.\ Mandelzweig, {\it Quasilinearization Method and  
Summation of the WKB Series}, 2004, submitted for publication.}

\bibitem{In} {E.~L.\ Ince, {\it Ordinary Differential Equations},
Dover Publications, New York, 1956.}

\bibitem{Lan} {R.~E.\ Langer, Phys.\ Rev.\ {51}, 669 (1937);
C.~M.\ Bender and S.~A.\ Orszag, {\it Advanced Mathematical
Methods for Scientists and Engineers I}, Springer-Verlag, New
York, 1999.}

\bibitem{ftn} {This form is based on a linear potential interpolation
near turning points from which the Airy function arises.}

\bibitem{M}
{J.~Malenfant, Phys.\ Rev.\ D{38}, 3295 (1988).}

\bibitem{S}
{T.~C.\ Scott, J.~Shertzer and R.~A.\ Moore, Phys.\ Rev.\
A{45}, 4393 (1992).}

\bibitem{Lee}
{R.~Friedberg, T.~D.\ Lee, Ann.\ Phys.\ {\bf 308}, 263 (2003).}

\bibitem{BB}
N.~E.~Jannik Bjerrum-Bohr, J.\ Math.\ Phys.\ 41, 2515 (2000).

\bibitem{Park}
D.~Park, Introduction to quantum theory, McGraw-Hill, New York 1964, p.\ 102.
\end{references}
\end{document}